\documentclass[12pt,letterpaper]{article}
\pdfoutput=1
\usepackage{jheppub}
\usepackage{amsfonts, amsthm}
\usepackage[english]{babel}
\usepackage[utf8]{inputenc}
\usepackage{slashed}
\hypersetup{unicode}

\newcommand{\eq}{\begin{equation}}
\newcommand{\feq}{\end{equation}}
\newcommand{\eqn}{\begin{eqnarray}}
\newcommand{\feqn}{\end{eqnarray}}

\newcommand{\ma}[1]{\mbox{$\mathcal{#1}$}}

\newcommand{\I}{\ma{I}}

\title{Rotating black holes in an expanding universe from fake supergravity}
 
\author{Samuele Chimento}
\author{and Dietmar Klemm}
\affiliation{Dipartimento di Fisica, Universit\`a di Milano, and \\
INFN, Sezione di Milano, \\
Via Celoria 16, 20133 Milano, Italy.
}
\emailAdd{samuele.chimento@mi.infn.it}
\emailAdd{dietmar.klemm@mi.infn.it}
\preprint{IFUM-1026-FT}

\abstract{Using the recipe of arXiv:0902.4814, where all fake supersymmetric backgrounds of
matter-coupled fake $N=2$, $d=4$ gauged supergravity were classified, we construct
dynamical rotating black holes in an expanding FLRW universe. This is done for two different
prepotentials that are both truncations of the stu model and correspond to just one vector multiplet.
In this scenario, the cosmic expansion is driven by two $\text{U}(1)$ gauge fields and by a complex
scalar that rolls down its potential. Generically, the solutions of arXiv:0902.4814 are fibrations over
a Gauduchon-Tod base space, and we make three different choices for this base, namely flat space,
the three-sphere and the Berger sphere. In the first two cases, the black holes are determined by
harmonic functions on the base, while in the last case they obey a deformed Laplace equation
that contains the squashing parameter of the Berger sphere. This is the generalization to a
cosmological context of the usual recipe in ungauged supergravity, where black holes are given in terms
of harmonic functions on three-dimensional Euclidean space. The constructed solutions may be
instrumental in addressing analytically questions like black hole collisions and violation of cosmic
censorship.
}


\begin{document}
\maketitle
\flushbottom

\section{Introduction}
Black holes are the natural test ground for quantum gravity. Much of the current knowledge on quantum effects in strong 
gravitational fields indeed comes from the study of stationary black holes. However many interesting open questions, such as
the validity of the cosmic censorship conjecture or what happens when black holes collide, are dynamical in nature and thus
require the study of time-dependent black hole solutions.

One well-known such solution is the McVittie spacetime \cite{McVittie:1933zz}, whose interpretation as a black hole, or a 
mass particle, in an FLRW universe has been the subject of some controversy in the literature 
\cite{Nolan:1998xs,Nolan:1999kk,Kaloper:2010ec}. Another example, which however violates the energy conditions, was 
constructed by Sultana and Dyer \cite{Sultana:2005tp} using conformal methods.

Kastor and Traschen (KT) \cite{Kastor:1992nn} obtained a solution representing an arbitrary number of
electrically charged black holes, with charge equal to the mass, in a de~Sitter universe. This solution
allows an analytical discussion of black hole collisions and of the issue whether such processes lead to
a violation of cosmic censorship \cite{Kastor:1992nn,Brill:1993tm}.
The KT solution is a time-dependent generalization of the Majumdar-Papapetrou 
spacetime \cite{Majumdar:1947eu,Papapetrou:1947xx}, which describes maximally charged Reissner-Nordström black holes in 
static equilibrium in an asymptotically flat space. The MP solution is supersymmetric, and the existence of a Killing spinor, 
satisfying a first order differential equation, explains why one can take arbitrary superpositions of black holes despite the 
high non-linearity of Einstein's equations. Supersymmetry however is only compatible with a negative or vanishing 
cosmological constant, thus no true Killing spinor can exist in a theory with positive cosmological constant. It was shown in \cite{Kastor:1993mj} that the KT solution admits instead a \emph{fake} Killing spinor, i.e.,
a solution of first order equations which are related to the Killing spinor equations of supergravity but do
not come from an underlying supersymmetry.

Maeda, Ohta and Uzawa (MOU) obtained four- and five-dimensional black holes in an FLRW universe
filled with stiff matter from the compactification of higher dimensional intersecting brane
solutions \cite{Maeda:2009zi}. In \cite{Gibbons:2009dr} 
Gibbons and Maeda presented a class of spacetimes interpolating between the KT and the four-dimensional MOU black holes as 
solutions to a theory with a Liouville-type scalar potential, later generalized to arbitrary dimension and further analyzed in \cite{Maeda:2010ja}. In \cite{Chimento:2012mg} the four-dimensional case was
generalized to a scalar potential given by 
a sum of exponentials and the black holes were shown to admit a fake Killing spinor, explaining the
superposition principle observed in the solution.

Only a few time-dependent rotating black hole solutions are known. A spinning generalization of the KT solution in a 
string-inspired theory was given by Shiromizu in \cite{Shiromizu:1999xj}. Five-dimensional multi-centered rotating charged 
de~Sitter black holes were constructed in \cite{Klemm:2000vn,Klemm:2000gh}. A rotating generalization of the five-dimensional 
MOU solution was obtained in \cite{Nozawa:2010zg} by solving fake Killing spinor equations.

In this paper we will use the classification of all the fake supersymmetric solutions of
Wick-rotated\footnote{In this context, by `Wick rotation' we mean $g\to ig$, where $g$ denotes the
coupling constant.}
$N=2$, $d=4$ gauged supergravity coupled to (non)abelian vector multiplets given in
\cite{Meessen:2009ma}\footnote{For a classification without matter coupling (pure fake $N=2$, $d=4$
gauged supergravity) see \cite{Gutowski:2009vb}.}
to build explicit time-dependent black hole solutions. We will restrict ourselves to the case of a single
abelian vector multiplet, corresponding to a theory
with two $U(1)$ gauge fields and a single complex scalar field. Unlike what we did in \cite{Chimento:2012mg}, we will not 
require the scalar to be real (or equivalently imaginary). This will allow us to obtain solutions with rotation and 
NUT-charge that are generalizations of a subclass of those in \cite{Chimento:2012mg}. For one choice of the prepotential 
defining the theory, these can be written in terms of two complex harmonic functions in a form similar to the IWP class of 
metrics \cite{Israel:1972vx,Perjes:1971gv}, of which they are generalizations.
We will also present solutions whose spatial slices have non-flat geometry. If the three-dimensional base space is spherical 
the solutions are given in terms of functions that are harmonics on the three-sphere. 

The paper is organized as follows. In section \ref{fake_sugra} we briefly review fake $N=2$, $d=4$ gauged supergravity
coupled to abelian vector multiplets and present the recipe of \cite{Meessen:2009ma} to construct fake supersymmetric
solutions. In section \ref{basespace} we consider three different geometries for the three-dimensional base space and obtain
some results that are independent of the specific theory (i.e., of the prepotential) under consideration.
We also show, for flat or spherical
geometry, how to obtain multi-centered solutions. In sections \ref{sec:mod1} and
\ref{sec:mod2} we obtain explicit solutions for two different choices of the prepotential. 
In section \ref{conclusion} we conclude with some final remarks.

\section{Fake \texorpdfstring{$N=2$, $d=4$}{N=2, d=4} gauged supergravity\label{fake_sugra}}
\label{fake-sugra}

\subsection{Special geometry}
In $N=2$, $d=4$ supergravity coupled to $n_V$ vector multiplets, the complex scalars of the multiplets
parametrize an $n_V$-dimensional K\"ahler-Hodge manifold, which is the base of a symplectic bundle
with the covariantly holomorphic sections\footnote{Here and in what follows we use the conventions
of \cite{Meessen:2009ma}.}
\begin{equation}
 \mathcal{V}=\left(\begin{array}{c}
                   \mathcal{L}^\Lambda\\
                   \mathcal{M}_\Lambda
                  \end{array}\right), \qquad \mathcal{D}_{\bar \imath}\mathcal{V}\equiv\partial_{\bar \imath}\mathcal{V}-\frac{1}{2}\left(\partial_{\bar \imath}\mathcal{K}\right)\mathcal{V}=0\,,
\end{equation}
obeying the constraint
\begin{equation}
 \left\langle\mathcal{V},\bar{\mathcal{V}}\right\rangle\equiv\bar{\mathcal{L}}^\Lambda\mathcal{M}_\Lambda-\mathcal{L}^\Lambda\bar{\mathcal{M}}_\Lambda=-i\,, \label{eq:sympcond}
\end{equation}
where $\mathcal{K}$ is the K\"ahler potential.
We also introduce the explicitly holomorphic section
\begin{equation}
 \Omega\equiv e^{-\mathcal{K}/2}\mathcal{V}\equiv\left(\begin{array}{c}
						  \mathcal{\chi}^\Lambda\\
						  \mathcal{F}_\Lambda
						  \end{array}\right)\,.
\end{equation}
If the theory is defined by a prepotential $\mathcal{F}(\chi)$, then $\mathcal{F}_\Lambda=\partial_\Lambda \mathcal{F}$.
In terms of the section $\Omega$ the constraint (\ref{eq:sympcond}) becomes
\begin{equation}
 \left\langle\Omega,\bar{\Omega}\right\rangle\equiv\bar{\chi}^\Lambda\mathcal{F}_\Lambda-\chi^\Lambda\bar{\mathcal{F}}_\Lambda=-i e^{-\mathcal{K}}.\label{eq:sympcond2}
\end{equation}
The couplings of the vectors to the scalars are determined by the matrix $\mathcal{N}$, defined by the 
relations
\begin{equation}
 \mathcal{M}_\Lambda = \mathcal{N}_{\Lambda\Sigma}\,\mathcal{L}^\Sigma, 
 \qquad \mathcal{D}_{\bar \imath}\bar{\mathcal{M}}_\Lambda=\mathcal{N}_{\Lambda\Sigma}\,\mathcal{D}_{\bar \imath}\bar{\mathcal{L}}^\Sigma\,.
\end{equation}
In a theory with a prepotential, $\mathcal{N}$ is given by
\begin{equation}
 \mathcal{N}_{\Lambda\Sigma}=\bar{\mathcal{F}}_{\Lambda\Sigma}+2i\frac{\mathfrak{Im}(\mathcal{F})_{\Lambda\Lambda'}\chi^{\Lambda'}\mathfrak{Im}(\mathcal{F})_{\Sigma\Sigma'}\chi^{\Sigma'}}{\chi^{\Omega}\mathfrak{Im}(\mathcal{F})_{\Omega\Omega'}\chi^{\Omega'}}\,, \label{eq:nmatrix}
\end{equation}
where $\mathcal{F}_{\Lambda\Sigma}=\partial_\Lambda\partial_\Sigma \mathcal{F}$.

The bosonic Lagrangian in the case of abelian vector multiplets, and with Fayet-Iliopoulos (FI) gauging
of a $\text{U}(1)$ R-symmetry subgroup, takes the form
\begin{align}
 e^{-1}\mathcal{L}_{\text{bos}}=&\,R+2\mathcal{G}_{i\bar\jmath}\partial_a Z^i\partial^a \bar Z^{\bar\jmath}-V
                                                          \nonumber\\
                         &+2\mathfrak{Im}(\mathcal{N})_{\Lambda\Sigma}F^\Lambda_{ab}F^{\Sigma ab}-2\mathfrak{Re}(\mathcal{N})_{\Lambda\Sigma}F^\Lambda_{ab}\star F^{\Sigma ab}\,,\label{eq:general_lagrangian}
\end{align}
with the scalar potential
\begin{equation}
 V=-\frac{g^2}{2}\left[4\left|C_\Lambda\mathcal{L}^\Lambda\right|^2+\frac{1}{2}\mathfrak{Im}(\mathcal{N})^{-1|\Lambda\Sigma}C_\Lambda C_\Sigma\right].\label{eq:scalar_pot}
\end{equation}
Here $g$ denotes the gauge coupling constant, and the FI parameters $C_\Lambda$ determine the
linear combination $C_\Lambda A^\Lambda$ that is used to gauge the $\text{U}(1)$.
Since the matrix $\mathfrak{Im}(\mathcal{N})_{\Lambda\Sigma}$ appears in the kinetic term of the vector fields,
it must be negative definite and thus invertible. It can therefore be used as a `metric' to raise and lower
$\Lambda,\Sigma,\dots$ indices.

\subsection{Fake Killing spinors}

If we perform a Wick rotation on the gauge coupling constant, $g\rightarrow ig$, we obtain a new, non-supersymmetric theory with $V\rightarrow -V$ and a gauged
$\mathbb{R}$-symmetry\footnote{Note that the resulting theory is different from the so-called
de~Sitter supergravities \cite{Pilch:1984aw}. To get the latter, one also takes $A_{\mu}\to iA_{\mu}$,
which leads to gauge field kinetic terms with the wrong sign, and thus to ghosts. In the theory
considered here, the kinetic terms of the gauge fields come with the correct sign. We thank P.~Meessen
for clarifying discussions on this point.}.
The Killing spinor equations, coming 
from the vanishing of the fermionic supersymmetry variations, become
\begin{align}
 \mathbb{D}_a\epsilon_I&=\left[-2i\mathcal{L}_\Lambda F^{\Lambda +}_{ab}\gamma^b-\frac{ig}{4}C_\Lambda\mathcal{L}^\Lambda\gamma_a\right]\varepsilon_{IJ}\epsilon^J\,,\nonumber\\
 i\slashed{\partial}Z^i \epsilon^I&=\left[\bar f^i_\Lambda\slashed{F}^{\Lambda +}-\frac{g}{2}C_\Lambda\bar f^{i\Lambda}\right]\varepsilon^{IJ}\epsilon_J\,,
\end{align}
where
\[
\mathbb{D}_a\epsilon_I\equiv\left(\nabla_a +\frac{i}{2}\mathcal{Q}_a-\frac{g}{2}C_\Lambda
A_a^\Lambda\right)\epsilon_I\,,
\]
$\mathcal{Q}_a=(2i)^{-1}\left(\partial_a Z^i\partial_i\mathcal{K}-\partial_a \bar Z^{\bar\imath}\partial_{\bar\imath}\mathcal{K}\right)$ is 
the gauge field of the K\"ahler $\text{U}(1)$, and $f_i^\Lambda\equiv \mathcal{D}_i \mathcal{L}^\Lambda=\left(\partial_i+\frac{1}{2}\partial_i\mathcal{K}\right)\mathcal{L}^\Lambda$.

Since these equations do not come from supersymmetry, they are called fake Killing spinor equations, and solutions for which
they are satisfied are known as fake supersymmetric.

From the fake Killing spinors one can construct the bilinears
\begin{equation}
X=\frac{1}{2} \varepsilon^{IJ}\bar\epsilon_I\epsilon_J\,, \qquad
V_a=i\bar\epsilon^I\gamma_a\epsilon_I\,, \qquad
V_a^x=i(\sigma^x)_I^{\phantom{I}J}\bar\epsilon^I\gamma_a\epsilon_J\,,\label{eq:bilinears}
\end{equation}
and the real symplectic sections of K\"ahler weight zero
\begin{equation}
\mathcal{R}\equiv \mathfrak{Re}(\mathcal{V}/X)\,, \qquad
\mathcal{I}\equiv \mathfrak{Im}(\mathcal{V}/X)\,. \label{eq:risections}
\end{equation}

\subsection{Fake supersymmetric solutions}
In \cite{Meessen:2009ma}, Meessen and Palomo-Lozano presented a general method to obtain fake
supersymmetric solutions to fake $N=2$, $d=4$ gauged supergravity coupled to nonabelian vector
multiplets. We will restrict ourselves here to the case of just abelian multiplets and FI gauging.
We will also consider only the timelike case of \cite{Meessen:2009ma}, which means
that we take the norm of $V$ defined in (\ref{eq:bilinears}) to be positive.
With these restrictions, the fake supersymmetric solutions always assume the form \cite{Meessen:2009ma}
\begin{align}
 ds^2&=2\left|X\right|^2(d\tau+\omega)^2-\frac{1}{2\left|X\right|^2}h_{mn}dy^mdy^n\,, \\
 A^\Lambda&=-\frac{1}{2}\mathcal{R}^\Lambda V + \tilde{A}^\Lambda_m dy^m\,, \\
 Z^\Lambda&=\frac{\mathcal{L}^\Lambda}{\mathcal{L}^0}=\frac{\mathcal{R}^\Lambda+i\mathcal{I}^\Lambda}{\mathcal{R}^0+i\mathcal{I}^0}\,,
\end{align}
where $V=2\sqrt{2}\left|X\right|^2(d\tau+\omega)$, $\omega=\omega_m dy^m$ is a 1-form which can in
general depend on $\tau$, and $h$ is the metric on a three-dimensional Gauduchon-Tod \cite{1998JGP....25..291G}
base space. In particular there must exist a dreibein $W^x$ for $h$ satisfying
\begin{equation}
 dW^x=gC_\Lambda \tilde{A}^\Lambda \wedge W^x +\frac{g}{2\sqrt{2}} C_\Lambda {\mathcal{I}}^\Lambda \varepsilon^{xyz} W^y\wedge W^z.\label{eq:gt}
\end{equation}
Furthermore the following equations must hold:
\begin{align}
 &\omega=gC_\Lambda\tilde A^\Lambda\tau+\tilde\omega\,, \label{eq:omegataudep}\\
 &\tilde{F}^\Lambda_{xy}=-\frac{1}{\sqrt{2}}\varepsilon^{xyz}\tilde{\mathbb{D}}_z \mathcal{I}^\Lambda\,,
 \label{eq:bogo}\\
 &\partial_\tau \mathcal{I}^\Lambda=0\,, \qquad \partial_\tau \mathcal{I}_\Lambda=-\frac g{2\sqrt2}
 C_\Lambda\,, \label{eq:dertaui}\\
 &\tilde{\mathbb{D}}_x^2\tilde{\mathcal{I}}_\Lambda-\left(\tilde{\mathbb{D}}_x \tilde \omega_x\right)
 \partial_\tau\mathcal{I}_\Lambda =0\,, \label{eq:omegai1}\\
 &\tilde{\mathbb{D}}\,\tilde \omega=\varepsilon^{xyz}\left\langle\tilde{\mathcal{I}}\right.\left|\partial_x
 \tilde{\mathcal{I}}-\tilde\omega_x\partial_\tau\mathcal{I}\right\rangle W^y\wedge W^z\,, \label{eq:omegai2}
\end{align}
with
\begin{gather}
\tilde{F}^\Lambda\equiv d \tilde{A}^\Lambda\,, \qquad \tilde\omega\equiv\omega|_{\tau=0}\,, \qquad
\tilde{\mathcal{I}}\equiv\mathcal{I}|_{\tau=0}\,, \\
\tilde{\mathbb{D}}_m \mathcal{I}\equiv \partial_m\mathcal{I}+gC_\Lambda\tilde A^\Lambda_m\mathcal{I}\,,
\qquad \tilde{\mathbb{D}}_x \mathcal{I}\equiv W_x^m\tilde{\mathbb{D}}_m \mathcal{I}\,.
\end{gather}
To obtain a specific solution we will then have to take the following steps:
\begin{enumerate}
 \item Choose the number of vector multiplets, the real constants $C_\Lambda$ and the prepotential
 $\cal F$. This completely determines the bosonic action and permits to
 derive the dependence of the 
 $\mathcal{R}$'s from the $\mathcal{I}$'s, the so-called \emph{stabilization equations}.
 \item Choose a three-dimensional Gauduchon-Tod base space, that is, choose a solution 
 $(W^x,C_\Lambda\tilde A^\Lambda,C_\Lambda \mathcal{I}^\Lambda)$ of equation (\ref{eq:gt}).
 \item Determine the $\mathcal{I}^\Lambda$'s and the $\tilde{A}^\Lambda$'s that respect the choices of points
 1 and 2 and at the same time satisfy equation (\ref{eq:bogo}).
 \item Determine the $\mathcal{I}_\Lambda$'s and $\tilde\omega$ from (\ref{eq:dertaui}) and the coupled
  equations (\ref{eq:omegai1}) and (\ref{eq:omegai2}).
 \item Solve the stabilization equations to find the $\mathcal{R}$'s and finally write down the metric and the
 other fields of the solution using (\ref{eq:omegataudep}) and
 $1/\left|X\right|^2=2\left\langle\mathcal{R}|\mathcal{I}\right\rangle$.
\end{enumerate}

In the next sections, we will use this procedure to find some solutions to theories with one vector multiplet, so
that there will be only one physical scalar $Z^1\equiv Z$.

\section{Choice of base space\label{basespace}}

\subsection{Flat space\label{baseflat}}

The simplest solution of eq. (\ref{eq:gt}) is three-dimensional flat space, with
\begin{equation}
 W^x_m=\delta^x_m,\qquad C_\Lambda\tilde A^\Lambda=C_\Lambda \mathcal{I}^\Lambda=0\,.
\end{equation}
With this choice for the base space we don't need to distinguish between $x,y,z\dots$ and lower
$m,n,p,\dots$ indices. 

If $C_0=C_1=0$, $C_\Lambda \mathcal{I}^\Lambda=0$ is automatically satisfied and the section $\ma{I}$
is time-independent.
Using equation (\ref{eq:bogo}) and the Bianchi identity $d\tilde F^\Lambda=0$ it can be seen that the $\ma{I}^\Lambda$ must be harmonic,
\begin{equation}
 \ma{I}^0\equiv\sqrt{2}H^0\,,\qquad\ma{I}^1\equiv\sqrt{2}H^1\,.
\end{equation}
Moreover, (\ref{eq:omegai1}) implies that the $\ma{I}_\Lambda$ are harmonic as well,
\begin{equation}
 \ma{I}_0\equiv\frac{H_0}{2\sqrt{2}}\,,\qquad\ma{I}_1\equiv\frac{H_1}{2\sqrt{2}}\,.
\end{equation}
Equ.~(\ref{eq:omegai2}) becomes
\begin{equation}
 d\tilde\omega=\star_3 \left( H_0dH^0+H_1dH^1- H^0dH_0-H^1dH_1\right)\label{eq:flat_ungauged_domega}.
\end{equation}
If at least one of the $C_\Lambda$ is nonzero, e.g.~$C_1\neq 0$, $C_\Lambda \mathcal{I}^\Lambda=0$ implies $\ma{I}^1=-\frac{C_0}{C_1} \ma{I}^0$.
Then, (\ref{eq:bogo}) and the Bianchi identity $d\tilde F^0=0$ yield
\begin{equation}
 \ma{I}^0=\sqrt{2}H_{\text{im}}\,,\qquad\ma{I}^1=-\sqrt{2}\,\frac{C_0}{C_1}H_{\text{im}}\,,
\end{equation}
where $H_{\text{im}}$ is a time-independent harmonic function\footnote{Since $H_{\text{im}}$ is related
to the imaginary part $\ma{I}^\Lambda$, the label `$\text{im}$' stands for `imaginary'.}.

(\ref{eq:omegai1}) together with (\ref{eq:dertaui}) implies that the time-independent combination 
$\ma{I}_0-\frac{C_0}{C_1}\ma{I}_1$ is harmonic. It proves convenient to express this defining
\begin{equation}
 \I_0\equiv\frac{C_0}{C_1}\left( \I_1 -\frac{1}{2\sqrt{2}}H_1\right)+\frac{1}{2\sqrt{2}}H_0\,,
\end{equation}
with $H_0$, $H_1$ harmonic functions independent of $\tau$.
Since there are no further constraints on $\tilde \I_1$, the $\I_\Lambda$ can be written as
\begin{equation}
 \I_1=\frac{1}{2\sqrt{2}}\left( \frac{\tau}{t_1}+f\right)\,,\qquad \I_0=\frac{1}{2\sqrt{2}}\left[ \frac{\tau}{t_0}+H_0+\frac{t_1}{t_0}(f-H_1)\right]\,,
\end{equation}
where $t_\Lambda\equiv-(gC_\Lambda)^{-1}$ and $f$ is a generic function of the spatial coordinates.

Equ.~(\ref{eq:omegai2}) becomes
\begin{equation}
 d\tilde\omega=\star_3\left[ \left( H_0-\frac{t_1}{t_0} H_1\right)dH_{\text{im}} - H_{\text{im}}
d\left( H_0-\frac{t_1}{t_0}H_1 \right) \right]\,, \label{eq:flatdomega}
\end{equation}
and from (\ref{eq:omegai1}) one gets
\begin{equation}
 \partial_p\tilde\omega_p=t_1\partial_p\partial_p f\,.\label{eq:flatdivomega}
\end{equation}
It is always possible to set $f$ to zero with a shift in the time coordinate, $\tau=t-t_1 f+t_1 H_1$, and
replacing $\tilde \omega$ by $\hat \omega=\tilde \omega-t_1 df+t_1dH_1$, such that
\begin{gather}
 \I_1=\frac{1}{2\sqrt{2}}\left( \frac{t}{t_1}+H_1\right)\,,\qquad \I_0=\frac{1}{2\sqrt{2}}\left( \frac{t}{t_0}+H_0\right)\,, \nonumber\\
 d\hat\omega=\star_3\left[ \left( H_0-\frac{t_1}{t_0} H_1\right)dH_{\text{im}} - H_{\text{im}}
d\left( H_0-\frac{t_1}{t_0} H_1 \right) \right]\,,\qquad\partial_p\hat\omega_p=0\,, \label{eq:flat}\\
d\tau+\tilde \omega=dt+\hat \omega\,. \nonumber
\end{gather}

An explicit choice for the harmonic functions, best expressed in Boyer-Lindquist coordinates $(r,\theta,\phi)$ with
$x+iy=\sqrt{r^2+a^2}\sin\theta e^{i\varphi}$ and $z=r\cos\theta$, is
\begin{equation}
 H=k+q\,\mathfrak{Re}\left( V \right)+Q\,\mathfrak{Im}\left( V \right)\,,\label{eq:flat_harm}
\end{equation}
with
\begin{equation}
 V=\frac{1}{r-ia \cos\theta}\,.
\end{equation}
If all the harmonics have this form, \eqref{eq:flat} is solved by
\begin{equation}
 \hat\omega=\frac{1}{\Sigma}\left[-\frac12 a\sin^2\theta\left(2\,\widehat{k Q}\,r+\widehat{q Q} \right)+\widehat{kq}\,(r^2+a^2)\cos\theta\right]d\varphi\,,\label{eq:flat_omega}
\end{equation}
where
\begin{equation}
 \Sigma=r^2+a^2\cos^2\theta\,,\qquad \widehat{x y}= \tilde x y_{\text{im}} - x_{\text{im}}
\tilde y\,,\qquad \tilde x= x_0-\frac{t_1}{t_0}x_1\,.
\end{equation}

This choice is also suitable to be generalized to the multi-centered case. To this end, define
\begin{equation}
 V(\vec x,a)=\frac{1}{\sqrt{x^2+y^2+(z-i a)^2}}\,,
\end{equation}
and consider harmonic functions of the form
\begin{equation}
 H=k+\sum_I\left( q_I \mathfrak{Re}(V_I)+Q_I\mathfrak{Im}(V_I) \right)\,,
\end{equation}
with $V_I\equiv V(\vec x-\vec x_I,a_I)$, where $\vec x_I$ is an arbitrary point in $\mathbb R^3$ and the parameter $a_I$ 
in general depends on $I$. As long as the charges are taken to satisfy ${q_{\text{im}}}_I=\alpha\,\tilde q_I$,
${Q_{\text{im}}}_I=\alpha\, \tilde Q_I$  for every $I$, with $\alpha$ independent of $I$,
(\ref{eq:flat}) reduces to
\begin{equation}
 d\hat\omega=(\alpha\tilde k-k_{\text{im}})\star_3d\tilde H\,,
\end{equation}
where $\tilde H=H_0-t_1H_1/t_0$. $\hat\omega$ is thus given by a sum over $I$ of terms of the form
(\ref{eq:flat_omega}), with $\widehat{q Q}=0$. More explicitly, (\ref{eq:flat_omega}) with these charge
constraints can be written in Cartesian coordinates and generalized to  
\begin{align}
\hat\omega=&-2 (\alpha\tilde k-k_{\text{im}})\sum_I \left[\frac{\tilde Q_I \mathfrak{Re}(V_I)}{|\vec x-\vec x_I|^2+a_I^2+1/|V_I|^2}-\frac{\tilde q_I \mathfrak{Im}(V_I)}{|\vec x-\vec x_I|^2+a_I^2-1/|V_I|^2}\right]\cdot\nonumber\\
  &\phantom{aaaaaaaaaaaaaaaaaaaaaaaaaaaaaaaaaa}\cdot a_I\left[\left(x-x_I\right)dy-\left(y-y_I\right)dx\right]\,.
\end{align}

\subsection{Three-sphere\label{basesphere}}

Since Gauduchon-Tod spaces are actually conformal classes, it would be possible to take any conformally
flat three-dimensional manifold as a base space simply by applying a conformal transformation to the
quantities in section \ref{baseflat} with appropriate conformal weights, leading to a nonzero
$C_\Lambda\tilde A^\Lambda$. This would however result in the same four-dimensional
solutions expressed in different coordinates.

On the other hand there is a different Gauduchon-Tod structure that can be defined on the same conformal class, giving 
nonequivalent four-dimensional solutions. Start from a 3-sphere, with metric in the form
\begin{equation}
 ds_3^2=\frac14\left[ d\theta^2+\sin^2\theta\, d\varphi^2+\left(d\psi+\cos\theta \,d\varphi\right)^2 \right]\,,\label{eq:sphere_metric}
\end{equation}
and choose the dreibein
\begin{align}
 W^1&=\frac12\left( \sin\psi \,d\theta-\sin\theta \cos\psi \,d\varphi \right)\,, \nonumber\\
 W^2&=\frac12\left( \cos\psi \,d\theta+\sin\theta\sin\psi \,d\varphi \right)\,, \nonumber\\ 
 W^3&=\frac12\left( d\psi+\cos\theta \,d\varphi\right)\,, \label{eq:sphere_dreibein}
\end{align}
that obeys
\begin{equation}
 dW^x=-\varepsilon^{xyz}W^y\wedge W^z\,. \label{eq:sphereWdiff}
\end{equation}
Thus, equ.~(\ref{eq:gt}) is satisfied with
\begin{equation}
 C_\Lambda\tilde A^\Lambda = 0\,, \qquad C_\Lambda \mathcal{I}^\Lambda =
-\frac{2\sqrt{2}}{g}\,. \label{eq:spheregt}
\end{equation}
A useful consequence of (\ref{eq:sphereWdiff}) is that with this frame choice we have for the associated spin 
connection 
$\omega_{y\phantom{x}z}^{\phantom{y}x\phantom{z}}-\omega_{z\phantom{x}y}^{\phantom{z}x\phantom{y}}=2\,\varepsilon^{xyz}$, where
$\omega_{x\phantom{y}z}^{\phantom{x}y\phantom{z}}\equiv W_x^{\phantom{x}\mu}\omega_{\mu\phantom{y}z}^{\phantom{\mu}y\phantom{z}}$, as can easily be seen from Maurer-Cartan's first structure 
equation. This in particular implies that for a scalar function $f$ on the sphere
\begin{equation}
 \partial_x\partial_x f=\nabla_m\nabla^m f\,, \label{eq:harmonic_sphere}
\end{equation}
where $\nabla$ is the Levi-Civita connection associated with the metric (\ref{eq:sphere_metric}), and
\begin{equation}
\left[\partial_x,\partial_y\right]=2\,\varepsilon^{xyz}\partial_z\,. \label{eq:sphere_dercomm}
\end{equation}

From (\ref{eq:spheregt}) it is clear that the ungauged theory, $C_0=C_1=0$, is incompatible with this
GT-structure, hence at least one of the $C_\Lambda$ must be nonzero. If $C_1\neq 0$,
\eqref{eq:spheregt} gives $\I^1=2\sqrt{2}\,t_1-\frac{t_1}{t_0}\I^0$, where the $t_\Lambda$ were
defined in section \ref{baseflat}.
The Bianchi identity $d\tilde F^0=0$, using (\ref{eq:sphereWdiff}), immediately
implies $\varepsilon^{xyz}\partial_x\tilde F^0_{yz}=0$. Plugging in the expression for $\tilde F^0_{xy}$
given by (\ref{eq:bogo}) and using (\ref{eq:harmonic_sphere}) one concludes that $\I^0$ must be harmonic on the sphere,
\begin{equation}
 \I^0=\sqrt{2} H_{\text{im}}\,, \qquad\I^1=\sqrt{2}\,\left(2\,t_1-\frac{t_1}{t_0}H_{\text{im}}\right)\,.
\end{equation}
Equations (\ref{eq:dertaui}) and (\ref{eq:omegai1}) again imply that the combination 
$\I_0-\frac{t_1}{t_0}\I_1$ is harmonic on the base space,
\begin{equation}
 \I_0=\frac{t_1}{t_0}\left( \I_1 -\frac{1}{2\sqrt{2}}H_1\right)+\frac{1}{2\sqrt{2}}H_0\,,
\end{equation}
while no additional constraint is imposed on $\tilde\I_1$, so one has
\begin{equation}
 \I_1=\frac{1}{2\sqrt{2}}\left( \frac{\tau}{t_1} + f\right)\,,\qquad\I_0=\frac{1}{2\sqrt{2}}\left( \frac{\tau}{t_0} +H_0+\frac{t_1}{t_0}(f-H_1)\right)\,,
\end{equation}
where a generic function $f$ on $\text{S}^3$ was introduced. (\ref{eq:omegai2}) becomes
\begin{equation}
 d\tilde \omega=\star_3\left[\left( H_0-\frac{t_1}{t_0}H_1 \right)dH_{\text{im}} - H_{\text{im}}
d\left( H_0-\frac{t_1}{t_0}H_1 \right)-2\,t_1df+2\,\tilde\omega\right]\,,
\end{equation}
with $\partial_x\tilde\omega_x=t_1\partial_x\partial_xf$ due to \eqref{eq:omegai1}.
Setting as before $f=0$ by taking $\tau=t-t_1f+t_1H_1$ and $\tilde\omega=\hat\omega+t_1df-t_1dH_1$,
one gets 
\begin{equation}
 \I_0=\frac{1}{2\sqrt{2}}\left( \frac{t}{t_0} +H_0\right)\,,\qquad\I_1=\frac{1}{2\sqrt{2}}\left( \frac{t}{t_1} +H_1\right)\,,
\end{equation}
and $\hat \omega$ satisfies
\begin{equation}
 d\hat \omega=\star_3\left[\tilde H dH_{\text{im}} - H_{\text{im}}d\tilde H-2\,t_1dH_1+2\,\hat\omega\right]\,,\qquad\partial_x\hat\omega_x=\nabla^m\hat\omega_m=0\,,\label{eq:spheredomega}
\end{equation}
with $\tilde H\equiv H_0-\frac{t_1}{t_0}H_1$. If the harmonics are chosen such as to satisfy
$dH_{\text{im}}\wedge d\tilde H=0$, the simplest solution to these equations is
$\mathring \omega=\frac12H_{\text{im}}d\tilde H-\frac12\tilde H dH_{\text{im}}+t_1dH_1$, with 
$d\mathring\omega=0$, and all other solutions can be obtained by adding arbitrary solutions of $d\omega-2\star_3\omega=0$, 
which implies $\nabla^m\omega_m=0$; these are clearly independent of the choice of harmonic functions.

To make an explicit choice for $\hat\omega$ and the harmonics it is convenient to work with the usual hyperspherical coordinates,
\begin{equation}
 ds_{\text{S}^3}^2=d\Psi^2+\sin^2\Psi\left( d\Theta^2+\sin^2\Theta\, d\Phi^2 \right)\,.
\label{eq:sphere_stand_metric}
\end{equation}
In these coordinates the simplest nontrivial choice of harmonic function on $\text{S}^3$ is
\begin{equation}
 H=k+q\frac{\cos\Psi}{\sin\Psi}\,,\label{eq:sph_harm_choice}
\end{equation}
which is singular in the points $\Psi=0,\pi$. In a neighbourhood of the singularities the metric on
$\text{S}^3$ is well approximated by the flat metric in spherical coordinates with $\Psi$ playing the
role of a radial coordinate, and $H\sim k+\frac{q}{\Psi}$. If all the harmonics are chosen to be of the form
\eqref{eq:sph_harm_choice}, the minimal $\hat\omega$ becomes
\begin{equation}
 \mathring\omega=\frac12\frac{\tilde k q_{\text{im}}-k_{\text{im}}\tilde q-2q_1t_1}{\sin^2\Psi}d\Psi\,,
\end{equation}
which is the differential of a harmonic function and as such can be set to zero by a shift in the time coordinate and a redefinition of the harmonics $H_0$ and $H_1$. This is equivalent to taking
$\mathring\omega=0$ from the beginning by imposing the constraint
\begin{equation}
 \tilde k q_{\text{im}}-k_{\text{im}}\tilde q-2q_1t_1=0\,. \label{eq:sphere_harm_constraint}
\end{equation}
The equation $d\omega=2\star_3\omega$, together with (\ref{eq:sphereWdiff}) and
(\ref{eq:sphere_dercomm}), implies 
$\partial_x\partial_x\omega_y=-8\omega_y$, which means that the components of $\omega$ with respect to the dreibein $W^x$
are spherical harmonics on $\text{S}^3$ with eigenvalue $1-n^2=-8$. Using the well-known expressions for these spherical 
harmonics and rewriting the one-forms $W^x$ in the coordinates (\ref{eq:sphere_stand_metric}) it is possible to obtain
the most general solution for $\omega$ which is regular on the three-sphere.
The metric (\ref{eq:sphere_metric}) is obtained by considering $\text{S}^3$ embedded in $\mathbb{C}^2$, $|z_1|^2+|z_2|^2=1$, and
taking the parametrization
\begin{equation}
z_1 = \cos\frac{\theta}{2}\,e^{\frac i2(\varphi+\psi)}\,, \qquad z_2 = \sin\frac{\theta}{2}\,
e^{\frac i2(\varphi-\psi)}\,.
\end{equation}
Comparing this with the usual parametrization for $\text{S}^3$ in $\mathbb{R}^4$ one obtains in the
coordinates (\ref{eq:sphere_stand_metric}) the expressions
\begin{align}
 W^1=&-\sin\Theta\sin\Phi \,d\Psi+\sin\Psi(\sin\Psi\cos\Phi-\cos\Psi\cos\Theta\sin\Phi)\,d\Theta\nonumber\\
     &-\sin\Psi\sin\Theta(\cos\Psi\cos\Phi+\sin\Psi\cos\Theta\sin\Phi)\,d\Phi\,, \nonumber\\
 W^2=&\sin\Theta\cos\Phi \,d\Psi+\sin\Psi(\sin\Psi\sin\Phi+\cos\Psi\cos\Theta\cos\Phi)\,d\Theta\\
     &-\sin\Psi\sin\Theta(\cos\Psi\sin\Phi-\sin\Psi\cos\Theta\cos\Phi)\,d\Phi\,, \nonumber\\
 W^3=&\cos\Theta\,d\Psi-\sin\Psi\cos\Psi\sin\Theta\,d\Theta-\sin^2\Psi\sin^2\Theta\,d\Phi\,,\nonumber
\end{align}
and the most general regular $\omega$ is
\begin{align}
 \omega=&(a\cos\Phi-b\sin\Phi)(\sin\Theta\,d\Psi+\sin\Psi\cos\Psi\cos\Theta\,d\Theta-\sin^2\Psi\sin\Theta\cos\Theta\,d\Phi)\nonumber\\
        &-\sin\Psi(a\sin\Phi+b\cos\Phi)(\sin\Psi\,d\Theta+\cos\Psi\sin\Theta\,d\Phi)\nonumber\\
        &-c(\cos\Theta\,d\Psi-\sin\Psi\cos\Psi\sin\Theta\,d\Theta+\sin^2\Psi\sin^2\Theta\,d\Phi)\,,\label{eq:sphere_omega}
\end{align}
where $a$, $b$, and $c$ are constants.

It is also possible to construct multi-centered solutions by taking sums of harmonic functions with singularities in arbitrary 
points on the 3-sphere. Given the standard embedding of $\text{S}^3$ in $\mathbb{R}^4$, the harmonic function 
$\frac{\cos\Psi}{\sin\Psi}$ can be written as
\begin{equation}
 h=\frac{x_1}{\sqrt{1-x_1^2}}\,,
\end{equation}
and the analogous harmonic function with singularities in any couple of antipodal points can be simply obtained by a rotation in $\mathbb{R}^4$ sending the point $(1,0,0,0)$, corresponding to $\psi=0$, in
one of the new points. However in this case one has in general $d\mathring \omega\neq 0$, and in order
to reinstate $d\mathring \omega=0$ while keeping the possibility of having an arbitrary number of black
holes in arbitrary positions and with independent charges one has to impose
$q_{\text{im}}=\alpha\,\tilde q$ for each of them, where $\alpha$ is a proportionality constant.

\subsection{Berger sphere\label{baseberger}}

A more general Gauduchon-Tod space can be defined starting from the Berger sphere \cite{1998JGP....25..291G}, which is a 
squashed $\text{S}^3$ or an $\text{SU}(2)$ group manifold with an
$\text{SU}(2)\times\text{U}(1)$-invariant metric
\begin{equation}
 ds_3^2=d\theta^2+\sin^2\!\theta\, d\varphi^2+\cos^2\!\mu\left(d\psi+\cos\theta \,d\varphi\right)^2.\label{eq:berger_metric}
\end{equation}
Given the well-known expressions for the left-invariant 1-forms
\begin{displaymath}
\sigma_1^L = \sin\psi \,d\theta-\sin\theta \cos\psi \,d\varphi\,, \quad
\sigma_2^L = \cos\psi \,d\theta+\sin\theta\sin\psi \,d\varphi\,, \quad
\sigma_3^L = d\psi+\cos\theta \,d\varphi\,,
\end{displaymath}
and for the right-invariant 1-forms
\begin{displaymath}
\sigma_1^R = \sin\varphi \,d\theta-\sin\theta \cos\varphi \,d\psi\,, \quad
\sigma_2^R = \cos\varphi \,d\theta+\sin\theta\sin\varphi \,d\psi\,, \quad
\sigma_3^R = d\varphi+\cos\theta \,d\psi\,,
\end{displaymath}
one can define the dreibein \cite{Chave:1991pv}
\begin{align}
 W^1&=\cos\mu \,\sigma_1^R \pm \sin\mu\left( \cos\theta\, \sigma_2^R-\sin\theta\sin\varphi\,\sigma_3^R\right)\,, \nonumber\\
 W^2&=\cos\mu \,\sigma_2^R \mp \sin\mu\left( \cos\theta\, \sigma_1^R+\sin\theta\cos\varphi\,\sigma_3^R\right)\,, \nonumber\\
 W^3&=\cos\mu \,\sigma_3^R \pm \sin\mu\sin\theta\left( \sin\varphi\,\sigma_1^R+ \cos\varphi\,\sigma_2^R\right)\,, \label{eq:berger_dreibein}
\end{align}
that satisfies
\begin{equation}
dW^x = \pm\sin\mu\cos\mu\,\sigma_3^L\wedge W^x-\frac{\cos\mu}{2}\varepsilon^{xyz}W^y\wedge
W^z\,, \label{eq:bergerWdiff}
\end{equation}
so that equation (\ref{eq:gt}) is satisfied with
\begin{equation}
 C_\Lambda\tilde A^\Lambda=\pm\frac{\sin\mu\cos\mu}{g}\,\sigma_3^L,\qquad C_\Lambda \mathcal{I}^\Lambda=-\frac{\sqrt{2}}{g}\cos\mu\,. \label{eq:CI_berger}
\end{equation}
Using Maurer-Cartan's first structure equation it is possible to see that for a scalar function on the Berger sphere
\begin{gather}
 \partial_x\partial_x f\pm2 \sin\mu\cos\mu\,{\sigma_3^L}_x\partial_x f=\nabla_m\nabla^m f\,. \label{eq:harmonic_berger}
\end{gather} 
%  \nonumber\\
%  (\omega_{xzy}-\omega_{yzx})\,\sigma_z^R=\varepsilon_{xyw}W^w
% \end{gather}
Again at least one of the $C_\Lambda$ must be nonzero. If we assume $C_1\neq 0$,
\eqref{eq:CI_berger} yields $\I^1=\sqrt{2}\,t_1\cos\mu-\frac{t_1}{t_0}\I^0$, where the
$t_\Lambda$ are defined as before.

The Bianchi identity $d\tilde F^\Lambda=0$, using (\ref{eq:bergerWdiff}), 
implies 
\[
\varepsilon^{xyz}\left(\partial_x\pm2\sin\mu\cos\mu\,{\sigma_3^L}_x\right)\tilde F^\Lambda_{yz}=0\,.
\]
Substituting the expression for $F^\Lambda_{xy}$ given by (\ref{eq:bogo}) and using
(\ref{eq:harmonic_berger}) one gets for $K_{\text{im}}\equiv\frac{1}{\sqrt{2}}\I^0$:
\begin{equation}
\nabla_m\left[\nabla^m\pm\sin\mu\cos\mu\,{\sigma_3^L}^m\right]K_{\text{im}} =
\left[\nabla^m\pm\sin\mu\cos\mu\,{\sigma_3^L}^m\right]\nabla_m K_{\text{im}} = 0\,.\label{eq:berger_pseudoharm1}
\end{equation}
Eqns.~(\ref{eq:dertaui}) and (\ref{eq:omegai1}) imply that the combination 
$\tilde K\equiv2\sqrt{2}(\I_0-\frac{t_1}{t_0}\I_1)$ satisfies
\begin{equation}
 \left(\nabla_m\nabla^m -\sin^2\mu \right)\tilde K=0\,, \label{eq:berger_pseudoharm2}
\end{equation}
while no additional constraint is imposed on $\tilde\I_1$, so one has
\begin{equation}
 \I_1=\frac{1}{2\sqrt{2}}\left( \frac{\tau}{t_1} + f\right)\,,\qquad\I_0=\frac{1}{2\sqrt{2}}\left( \frac{\tau}{t_0} +\tilde K+\frac{t_1}{t_0}f\right)\,,
\end{equation}
where a generic function $f(\theta,\varphi,\psi)$ was introduced. (\ref{eq:omegai2}) becomes
\begin{equation}
 d\tilde \omega\pm\sin\mu\cos\mu\,\sigma_3^L\wedge\tilde\omega=\star_3\left[\tilde K dK_{\text{im}}
 - K_{\text{im}}d\tilde K-t_1\cos\mu\, df+\cos\mu\,\tilde\omega\right]\,,
\end{equation}
and from (\ref{eq:omegai1}) we get
\begin{equation}
\nabla_m \tilde\omega^m\mp \sin\mu\cos\mu\, {\sigma_3^L}_m\tilde\omega^m=t_1\left(\nabla_m\nabla^m -\sin^2\mu \right)f\,.
\end{equation}
It is possible to set $f=0$ by taking $\tau=t-t_1f+t_1K_1$ and 
$\tilde\omega=\hat\omega+t_1d(f-K_1)\pm\sin\mu\cos\mu\,\sigma_3^Lt_1 (f-K_1)$, where
$K_1(\theta,\varphi,\psi)$ satisfies (\ref{eq:berger_pseudoharm2}). In this way
\begin{equation}
 \I_0=\frac{1}{2\sqrt{2}}\left( \frac{t}{t_0} +K_0\right)\,,\qquad\I_1=\frac{1}{2\sqrt{2}}\left( \frac{t}{t_1} +K_1\right)\,,
\end{equation}
with $K_0\equiv\tilde K+\frac{t_1}{t_0}K_1$, and $\hat \omega$ satisfies
\begin{gather}
 d\hat \omega\pm\sin\mu\cos\mu\,\sigma_3^L\wedge\hat\omega=\star_3\left[\tilde K dK_{\text{im}}
-K_{\text{im}}d\tilde K-t_1\cos\mu\, dK_1+\cos\mu\,\hat\omega\right]\,,\nonumber\\
 \nabla^m\hat\omega_m\mp \sin\mu\cos\mu\, {\sigma_3^L}_m\hat\omega^m=0\,.\label{eq:berger_omega}
\end{gather}
There is no obvious way of finding solutions to the eqns.~(\ref{eq:berger_pseudoharm1}) and
(\ref{eq:berger_pseudoharm2}) that in the limit $\mu\rightarrow 0$ reduce to harmonic functions of the
form given in section \ref{basesphere}, which is what one would expect for black hole solutions. It is
however possible to consider simple solutions given by the trivial choices
\begin{equation}
K_0  =K_1 = 0\,, \qquad K_{\text{im}} = k_{\text{im}}\,, \qquad \hat\omega = 0\,, \label{eq:berger_trivial}
\end{equation}
with $k_{\text{im}}$ constant.

\section{The \texorpdfstring{$\mathcal{F}(\chi)=-\frac i4\chi^0\chi^1$}{ℱ(χ)=-i/4 χ⁰χ¹} model\label{sec:mod1}}

Given this prepotential, from (\ref{eq:sympcond2}) we can derive the K\"ahler potential
\begin{equation}
 e^{-\mathcal{K}}=\mathfrak{Re}(Z)\,,
\end{equation}
where we fixed $\left|\chi^0\right|=1$. The K\"ahler metric is then
\begin{equation}
\mathcal{G}=\partial_Z\partial_{\bar Z}\mathcal{K}=\frac14 \mathfrak{Re}(Z)^{-2}.
\end{equation}
From equation (\ref{eq:nmatrix}) one obtains
\begin{equation}
 \mathcal{N}=-\frac{i}{4}\left(\begin{array}{cc}
                    Z & 0                 \\
                    0 & \frac1Z
                   \end{array}
\right),
\end{equation}
and for the scalar potential (\ref{eq:scalar_pot}) one gets
\begin{equation}
V=g^2\left[\frac{C_0^2}{\mathfrak{Re}(Z)}+4 C_0 C_1
 +\frac{C_1^2}{\mathfrak{Re}(1/Z)} \right]\,. \label{eq:scalar_pot1}
\end{equation}
\eqref{eq:risections} leads to
\begin{equation}
 \mathcal{R}^0=-4\mathcal{I}_1\,, \qquad \mathcal{R}^1=-4 \mathcal{I}_0\,, \qquad
 \mathcal{R}_0=\frac14\mathcal{I}^1\,, \qquad \mathcal{R}_1=\frac14\mathcal{I}^0\,, \label{eq:mod1ir}
\end{equation}
as well as
\begin{equation}
 \frac{1}{2|X|^2}=\langle\mathcal{R}|\mathcal{I}\rangle=\frac12\,\mathcal{I}^0\mathcal{I}^1+8\,\mathcal{I}_0\mathcal{I}_1\,. \label{eq:mod1sympprod}
\end{equation}

\subsection{Flat base space}

Using the results of section \ref{baseflat}, one gets in the ungauged case from (\ref{eq:mod1sympprod})
\begin{equation}
 \frac{1}{2|X|^2}=H^0H^1+H_0H_1\,,
\end{equation}
and the solution takes the well-known form \cite{Behrndt:1997ny}
\begin{equation}
 ds^2=2|X|^2(d\tau+\tilde \omega)^2- \frac{1}{2|X|^2} d\vec y^{\,2}\,,  \qquad
Z=\frac{H_0-i H^1}{H_1-iH^0}\,, \label{eq:mod1_ungauged_sol}
\end{equation} 
\begin{displaymath}
F^0 = d\left( 2|X|^2 H_1(d\tau+\tilde \omega) \right)-\star_3dH^0\,, \qquad
F^1 = d\left( 2|X|^2 H_0(d\tau+\tilde \omega) \right)-\star_3dH^1\,,
\end{displaymath}
with $\tilde \omega$ satisfying (\ref{eq:flat_ungauged_domega}). In
the gauged case the solution can be written as
\begin{gather}
 ds^2=2|X|^2(dt+\hat \omega)^2- \frac{1}{2|X|^2} d\vec y^{\,2}\,,  \qquad
 Z = \frac{t/t_0+H_0+i t_1/t_0 H_{\text{im}}}{t/t_1+H_1-iH_{\text{im}}}\,, \label{eq:mod1_gauged_sol} \\
 F^0 = d\left[ 2|X|^2\left(\frac{t}{t_1}+H_1\right)(dt+\hat \omega) \right]-\star_3dH_{\text{im}}\,,
 \nonumber \\
 F^1 = d\left[ 2|X|^2\left(\frac{t}{t_0}+H_0\right)(dt+\hat\omega)\right]+\frac{t_1}{t_0}
 \star_3dH_{\text{im}}\,, \nonumber
\end{gather}
where
\begin{equation}
 \frac{1}{2|X|^2}=\left( \frac{t}{t_0} +H_0\right)\left( \frac{t}{t_1} +H_1\right)-\frac{t_1}{t_0}H_{\text{im}}^2\label{eq:mod1_gauged_met_fun}
\end{equation}
and $\hat\omega\equiv\tilde\omega-t_1df+t_1dH_1$ satisfies equ.~(\ref{eq:flat}).

Both solutions can also be rewritten in terms of two complex harmonic functions $\ma{H}_\Lambda$
as follows:
\begin{gather}
 ds^2=\frac{1}{\mathfrak{Re}(\ma{H}_0\bar{\ma{H}}_1)}(dt+\omega)^2 -  \mathfrak{Re}  (\ma{H}_0\bar{\ma{H}}_1)d\vec y^{\,2}\,,  \qquad
 Z = \frac{\ma{H}_0}{\ma{H}_1}\,, \label{eq:mod1_compl_sol} \\
 F^0 = d\left[ \frac{\mathfrak{Re}(\ma{H}_1)}{\mathfrak{Re}(\ma{H}_0\bar{\ma{H}}_1)}(dt+\omega) \right]+\star_3d\mathfrak{Im}(\ma{H}_1)\,,\nonumber \\
 F^1 = d\left[ \frac{\mathfrak{Re}(\ma{H}_0)}{\mathfrak{Re}(\ma{H}_0\bar{\ma{H}}_1)}(dt+\omega) \right]+\star_3d\mathfrak{Im}(\ma{H}_0)\,,\nonumber
\end{gather}
where $\omega$ is time-independent and satisfies
\begin{equation}
 d\omega= \star_3 \mathfrak{Im}\left( \ma{H}_0d\bar{\ma{H}}_1+\ma{H}_1d\bar{\ma{H}}_0 \right)\,.
\end{equation}
In the ungauged case, the only additional constraint on the complex harmonics is that they are
independent of time. In terms of the harmonics defined above they are given by
\begin{equation}
 \ma{H}_0=H_0-iH^1\,,\qquad \ma{H}_1=H_1-iH^0\,.
\end{equation}
In the gauged case the time dependence of the harmonics is completely determined by
$\partial_t \ma{H}_\Lambda=1/t_\Lambda$\footnote{Here one recognizes the substitution principle
originally put forward by Behrndt and Cveti\v{c} in \cite{Behrndt:2003cx}, which amounts to adding a
linear time dependence to the harmonic functions in a supersymmetric black hole of ungauged
$N=2$, $d=4$ supergravity.}.
In addition they must satisfy
$\mathfrak{Im}(\ma{H}_0)=-\frac{t_1}{t_0}\mathfrak{Im}(\ma{H}_1)$, and thus
\begin{equation}
 \ma{H}_0=\frac{t}{t_0}+H_0+i\frac{t_1}{t_0}H_{\text{im}}\,,\qquad \ma{H}_1=\frac{t}{t_1}+H_1-iH_{\text{im}}\,.
\end{equation}
In this case there is also the additional constraint $\partial_p\omega_p=0$.

Notice that (\ref{eq:mod1_compl_sol}) reduces to the Israel-Wilson-Perjés \cite{Perjes:1971gv,Israel:1972vx} solution for $\ma{H}_0=\ma{H}_1$. This means in particular that we can recover the Kerr-Newman solution with mass equal to the charge by 
taking
\begin{equation}
 \ma{H}_0=\ma{H}_1=1+qV\equiv 1+\frac{q}{r-ia \cos\theta}\,,\qquad \omega=
\frac{q a\sin^2\!\theta(2r+q)}{r^2+a^2\cos^2\!\theta}d\varphi\,,
\end{equation}
expressed in Boyer-Lindquist coordinates\footnote{One might ask whether the solution
\eqref{eq:mod1_compl_sol} has a minimal fake gauged supergravity limit. However, it is easy to
see that requiring the scalar $Z$ to be constant implies $t_\Lambda\to\infty$ with $t_1/t_0$ fixed, and
thus $g\to 0$, which brings us back to the ungauged case. This is consistent with the fact that (for
nonvanishing rotation) the Kerr-Newman-de~Sitter solution can never admit fake Killing spinors, as can
be seen by analytically continuing the
BPS condition (3.27) of \cite{AlonsoAlberca:2000cs} for the Carter-Pleba\'nski solution with
$\Lambda<0$, whose KNdS limit cannot be taken. We thank M.~Nozawa for pointing out this.}. 

This construction suggests the more general form (\ref{eq:flat_harm}) for the harmonics, with
$\omega$ given by (\ref{eq:flat_omega}). With these choices the gauged solution explicitly reads
\begin{align}
 ds^2=&\frac{\Sigma^2}{\Delta} dt^2+\frac{\Sigma}{\Delta}\left[-a\sin^2\!\theta\left(2\,\widehat{k Q}\,r+\widehat{q Q} \right)+2\,\widehat{kq}\,(r^2+a^2)\cos\theta\right] dt d\varphi\nonumber\\
      &-\frac{\Delta}{\Sigma (r^2+a^2)}dr^2-\frac{\Delta}{\Sigma}d\theta^2 \\
      &+\left[\frac1{4\Delta}\!\left[-a\sin^2\!\theta\left(2\,\widehat{k Q}\,r+\widehat{q Q} \right)\!+\!2\,\widehat{kq}\,(r^2+a^2)\cos\theta\right]^2\!\! - \frac{\Delta}{\Sigma^2}(r^2+a^2)\sin^2\!\theta
\right]d\varphi^2\,,
\nonumber
\end{align}
\begin{align}
    A^0=&\frac{\Sigma}{\Delta}\left(\Sigma(t/t_1+k_1)+q_1 r+Q_1 a \cos\theta\right) dt\nonumber\\
        &-\frac12 \left[\frac{\Sigma}{\Delta}\left(\Sigma(t/t_1+k_1)+q_1 r+Q_1 a \cos\theta\right)
          \left(2\widehat{k Q}r+\widehat{q Q} \right)-2 Q_{\text{im}}r\right]
          \frac{a\sin^2\!\theta}{\Sigma}d\varphi \nonumber \\
        &+\left[\frac{\Sigma}{\Delta}\left(\Sigma(t/t_1+k_1)+q_1 r+Q_1 a \cos\theta\right)
          \widehat{k q}-q_{\text{im}}\right]\frac{(r^2+a^2) \cos\theta}{\Sigma}d\varphi\,,
\end{align}
\begin{align}
    A^1=&\frac{\Sigma}{\Delta}\left(\Sigma(t/t_0+k_0)+q_0 r+Q_0 a \cos\theta\right) dt\nonumber\\
        &-\frac12 \left[\frac{\Sigma}{\Delta}\left(\Sigma(t/t_0+k_0)+q_0 r+Q_0 a \cos\theta\right)
          \left(2\widehat{k Q}r+\widehat{q Q} \right)+2\frac{t_1}{t_0} Q_{\text{im}}r\right]
          \frac{a \sin^2\!\theta}{\Sigma}d\varphi \nonumber\\
        &+\left[\frac{\Sigma}{\Delta}\left(\Sigma(t/t_0+k_0)+q_0 r+Q_0 a \cos\theta\right)\widehat{k q}+\frac{t_1}{t_0}q_{\text{im}}\right]\frac{(r^2+a^2) \cos\theta}{\Sigma}d\varphi\,,
\end{align}
\begin{equation}
    Z=\frac{\Sigma (t/t_0+k_0)+q_0 r+Q_0 a \cos\theta+i t_1/t_0(\Sigma k_{\text{im}}+q_{\text{im}}r+
        Q_{\text{im}}a\cos\theta)}{\Sigma (t/t_1+k_1)+q_1 r+Q_1 a \cos\theta-i(\Sigma k_{\text{im}}+
        q_{\text{im}}r+ Q_{\text{im}}a\cos\theta)}\,,
\end{equation}
where
\begin{align}
 \Delta=&\left[\Sigma\left(\frac{t}{t_0}+k_0\right)+q_0 r+Q_0 a \cos\theta\right]\left[\Sigma
               \left(\frac{t}{t_1}+k_1\right)+q_1 r+Q_1 a \cos\theta\right] \nonumber \\
             &-\frac{t_1}{t_0}\left[\Sigma k_{\text{im}}+q_{\text{im}}r+Q_{\text{im}}a\cos\theta\right]^2\,, \\
 \Sigma=&r^2+a^2\cos^2\!\theta\,, \qquad \widehat{x y} = \tilde x y_{\text{im}}-x_{\text{im}}\tilde y\,,
                 \qquad \tilde x= x_0-\frac{t_1}{t_0}x_1\,.
\end{align}
It can be seen from these expressions that the constant $\widehat{k q}$ in $\omega$ represents
essentially a NUT charge.

\subsection{Spherical base space}

Using the results of section \ref{basesphere}, the complete solution can be written in terms of harmonic
functions $H_{\text{im}}$, $H_0$, $H_1$ on $\text{S}^3$ and a time-independent one-form
$\hat\omega$ as
\begin{gather}
 ds^2=2|X|^2(dt+\hat \omega)^2- \frac{1}{2|X|^2} ds_{\text{S}^3}^2\,,  \nonumber\\
 F^0=d\left[ 2|X|^2 \left(  \frac{t}{t_1}+H_1\right)(dt+\hat \omega) \right]-\star_3dH_{\text{im}}\,,
 \nonumber \\
 F^1=d\left[ 2|X|^2 \left(  \frac{t}{t_0}+H_0\right)(dt+\hat \omega) \right]+\frac{t_1}{t_0}
 \star_3dH_{\text{im}}\,, \nonumber \\
 Z=\frac{t/t_0+H_0-i2\,t_1+i t_1H_{\text{im}}/t_0}{t/t_1+H_1-iH_{\text{im}}}\,, \label{eq:mod1_gauged_sol_spher}
\end{gather}
where
\begin{equation}
 \frac{1}{2|X|^2}=\left( \frac{t}{t_0} +H_0\right)\left( \frac{t}{t_1} +H_1\right)+H_{\text{im}}
 \left(2\,  t_1-\frac{t_1}{t_0}H_{\text{im}}\right)\,, \label{eq:mod1_gauged_met_fun_spher}
\end{equation}
and $\hat\omega$ satisfies (\ref{eq:spheredomega}). In particular the harmonics can be taken to be of
the form (\ref{eq:sph_harm_choice}), with $\hat\omega$ as in section \ref{basesphere}. The curvature
scalars $R$, $R_{\mu\nu}R^{\mu\nu}$ and $R_{\mu\nu\rho\sigma}R^{\mu\nu\rho\sigma}$ 
are singular for $\frac{1}{2|X|^2}=0$, but not in the points $\psi=0,\pi$ unless $q_0 q_1=\frac{t_1}{t_0} q_{\text{im}}^2$.

Note finally that the scalar field \eqref{eq:mod1_gauged_sol_spher} assumes the constant value
$Z=t_1/t_0$ (where the potential \eqref{eq:scalar_pot1} has an extremum\footnote{We assume $t_1/t_0>0$.}) if $t_0H_0=t_1H_1$ and $H_{\text{im}}=t_0$. In this case, $\tilde H=0$ and
$\mathring\omega=t_1dH_1$. If we take $\omega=0$ and define a new time coordinate $\tau$ by
$t+t_1H_1=t_0t_1\sinh\tau$, the metric becomes
\eq
ds^2 = t_0t_1\left[d\tau^2 - \cosh^2\tau ds^2_{\text{S}^3}\right]\,,
\feq
and the gauge field strengths $F^{\Lambda}$ vanish, so that the solution is dS$_4$. For $\omega\neq 0$,
one gets a deformation of dS$_4$ with nonzero $F^{\Lambda}$. This is what happens also in the `asymptotic' limit $\Psi\sim \pi/2$ of the solution with the explicit choice (\ref{eq:sph_harm_choice})
and with $t_0k_0=t_1k_1$, $k_{\text{im}}=t_0$. 

\subsection{Berger sphere}

For this base space, the results of section \ref{baseberger} imply that the complete solution can be
written in the form
\begin{align}
 ds^2&=2|X|^2(dt\pm\sin\mu\cos\mu\,\sigma_3^L\,t+\hat \omega)^2- \frac{1}{2|X|^2} ds_3^2\,,  \nonumber\\
 F^0&=d\left[ 2|X|^2 \left(  \frac{t}{t_1}+K_1\right)(dt\pm\sin\mu\cos\mu\,\sigma_3^L\,t+\hat \omega) \right]\nonumber\\
    &\qquad\qquad-\star_3[dK_{\text{im}}\pm\sin\mu\cos\mu\,\sigma_3^L\,K_{\text{im}}]\,,\nonumber \\
 F^1&=d\left[ 2|X|^2 \left(  \frac{t}{t_0}+K_0\right)(dt\pm\sin\mu\cos\mu\,\sigma_3^L\,t+\hat\omega)
\right]\nonumber\\
    &\qquad\qquad+\frac{t_1}{t_0}\star_3[dK_{\text{im}}\pm\sin\mu\cos\mu\,\sigma_3^L\,
 (K_{\text{im}}-t_0\cos\mu)]\,, \nonumber\\
 Z&=\frac{t/t_0+K_0-it_1\cos\mu+i t_1K_{\text{im}}/t_0}{t/t_1+K_1-iK_{\text{im}}}\,, \label{eq:mod1_gauged_sol_berg}
\end{align}
where
\begin{equation}
 \frac{1}{2|X|^2}=\left( \frac{t}{t_0} +K_0\right)\left( \frac{t}{t_1} +K_1\right)+
 K_{\text{im}}\left(t_1\cos\mu-\frac{t_1}{t_0}K_{\text{im}}\right)\,, \label{eq:mod1_gauged_met_fun_berg}
\end{equation}
the functions $K_0$ and $K_1$ satisfy (\ref{eq:berger_pseudoharm2}), $K_{\text{im}}$ satisfies (\ref{eq:berger_pseudoharm1}), and the time-independent one-form $\hat \omega$ is a solution of (\ref{eq:berger_omega}).

With the trivial choices (\ref{eq:berger_trivial}) the solution reduces to
\begin{gather}
 ds^2=\frac{t_0t_1}{t^2+\alpha_0\alpha_1}(dt\pm\sin\mu\cos\mu\,\sigma_3^L\,t)^2- \frac{t^2+\alpha_0\alpha_1}{t_0t_1}ds_3^2\,,  \nonumber\\
 A^\Lambda=\pm t_\Lambda\sin\mu\left(\frac{t^2 \cos\mu}{t^2+\alpha_0\alpha_1}-\frac{\alpha_\Lambda}{t_0 t_1} \right)\sigma_3^L\,,\quad Z=\frac{t_1}{t_0}\frac{t-i\alpha_1}{t-i\alpha_0}\,, 
\end{gather}
with
\begin{equation}
 \alpha_0=t_1 k_{\text{im}}\,, \qquad\alpha_1=t_0t_1\cos\mu-\alpha_0=t_0t_1\cos\mu-t_1 k_{\text{im}}\,.
\end{equation}
Imposing $\alpha_0=\alpha_1$, the scalar becomes constant and one obtains a solution of Einstein-Maxwell-de Sitter theory 
already found by Meessen \cite{Meessen:2010notes}. This can be seen as a deformation of dS$_4$, which is recovered for $\mu=0$.

\section{The \texorpdfstring{$\mathcal{F}(\chi)=-\frac18\frac{(\chi^1)^3}{\chi^0}$}{ℱ(χ)=-1/8 (χ¹)³/χ⁰} model\label{sec:mod2}}

Using (\ref{eq:sympcond2}) this prepotential leads to the K\"ahler potential
\begin{equation}
 e^{-\mathcal{K}}=\mathfrak{Im}(Z)^3\,,
\end{equation}
where we took $\left|\chi^0\right|=1$, and to the K\"ahler metric
\begin{equation}
\mathcal{G}=\partial_Z\partial_{\bar Z}\mathcal{K}=\frac34 \mathfrak{Im}(Z)^{-2}\,.
\end{equation}
The vectors' kinetic matrix is, according to equ.~(\ref{eq:nmatrix}),
\begin{equation}
 \mathcal{N}=\frac{1}{4}\left(\begin{array}{cc}
                    -Z\, \mathfrak{Re}(Z)^2-\frac i2 |Z|^2\,\mathfrak{Im}(Z)& \frac32 Z\, \mathfrak{Re}(Z)              \\
                    \frac32 Z \,\mathfrak{Re}(Z) & -3 Z+i \frac32\,\mathfrak{Im}(Z)
                   \end{array}
\right),
\end{equation}
and from (\ref{eq:scalar_pot}) one gets the scalar potential
\begin{equation}
V=\frac43\, g^2\frac{C_1^2}{\mathfrak{Im}(Z)}\,.
\end{equation}
It is worth noting that for the choice $C_1=0$ (and $C_0$ arbitrary) the potential vanishes (so-called
flat gauging), and the fake supersymmetric
solutions constructed here are also solutions to the equations of motion of the 
corresponding ungauged supergravity.

Requiring $\mathfrak{Re}(Z),\,\mathfrak{Im}(Z)\neq 0$ and $\langle\mathcal{R}|\mathcal{I}\rangle> 0$ the stabilization
equations give
 \begin{gather}
  \mathcal{R}^0=\frac{1}{2\,\ma{S}}\left[(\I^1)^3+4\,\I^0\I_1\I^1+4\,\I_0(\I^0)^2 \right]\,,\nonumber\\
  \mathcal{R}^1=-\frac{2}{9\,\ma{S}} \left[ 16\, \I^0(\I_1)^2+3\,\I_1(\I^1)^2-9\,\I_0\I^0\I^1\right]\,, \nonumber\\
  \mathcal{R}_0=\frac{2}{27\,\ma{S}}\left[16\,(\I_1)^3-27\,(\I_0)^2\I^0-27\,\I_0\I_1\I^1\right]\,,\nonumber\\
  \mathcal{R}_1=\frac{1}{6\,\ma{S}}\left[4\,(\I_1)^2\I^1-12\,\I_0\I^0\I_1-9\,\I_0(\I^1)^2\right]\,, \label{eq:mod2ir}
 \end{gather}
with
\begin{equation}
 \ma{S}\equiv\sqrt{-4(\I_0\I^0)^2+\frac{4}{3}(\I_1\I^1)^2+\frac{128}{27}\I^0(\I_1)^3-2\I_0(\I^1)^3-8\I_0\I^0\I_1\I^1}\,,
\end{equation}
and
\begin{equation}
 \frac{1}{2|X|^2}=\langle\mathcal{R}|\mathcal{I}\rangle=\ma{S}\,. \label{eq:mod2sympprod}
\end{equation}

\subsection{Flat base space}

Using again the results of section \ref{baseflat} the solution in the gauged case can be written in terms
of harmonic functions $H_0\,,H_1$ and $H_{\text{im}}$ and a time-independent one-form $\omega$ as
\begin{gather}
 ds^2=\ma{S}^{-1}(dt+\omega)^2- \ma{S} d\vec y^{\,2}\,, \qquad
 Z=\frac{\ma{T}^1+i t_1\ma{S}/t_0}{\ma{T}^0-i\ma{S}}\,, \label{eq:mod2_sol} \\
 F^0=d\left( \frac{H_{\text{im}}\ma{T}^0}{\ma{S}^2}(dt+ \omega) \right)-\star_3dH_{\text{im}}\,, \qquad
 F^1=d\left( \frac{H_{\text{im}}\ma{T}^1}{\ma{S}^2}(dt+ \omega) \right)+\frac{t_1}{t_0}\star_3
 dH_{\text{im}}\,, \nonumber
\end{gather}
with
\begin{gather}
 \ma{S}=\sqrt{H_{\text{im}}\ma{H}_0\left[\ma{T}^0+\left( \frac{t_1}{t_0} \right)^3{H_{\text{im}}}^2\right]+H_{\text{im}}\ma{H}_1\left[ \ma{T}^1-\frac{4}{27}{\ma{H}_1}^2\right]}\,, \nonumber\\
 \ma{T}^0=H_{\text{im}}\left[\left(\frac{t_1}{t_0}\right)^3 H_{\text{im}}-\ma{H}_0+
 \frac{t_1}{t_0}\ma{H}_1\right]\,, \nonumber\\
 \ma{T}^1=\frac{4}{9}{\ma{H}_1}^2+\frac{1}{3}\left( \frac{t_1}{t_0} \right)^2 H_{\text{im}}\ma{H}_1+
 \frac{t_1}{t_0}H_{\text{im}}\ma{H}_0\,, \nonumber\\
 \ma{H}_0=\frac{t}{t_0}+H_0\,,\qquad \ma{H}_1=\frac{t}{t_1}+H_1\,,
\end{gather}
while $\omega$ solves equ.~(\ref{eq:flat}).

In the case $C_0=0$ ($t_0\rightarrow\infty$) and with the convenient redefinitions
$H_1\rightarrow 3/2\, H_1$, $\tilde t_1=3/2\, t_1$ the solution simplifies to
\begin{gather}
 ds^2=\ma{S}^{-1}(dt+\omega)^2- \ma{S} d\vec y^{\,2}\,,  \qquad
 Z=-\frac{\ma{H}_1^2}{H_0 H_{\text{im}}+i\ma{S}}\,, \\
 F^0=-d\left( \frac{H_0 H_{\text{im}}^2}{\ma{S}^2}(dt+ \omega) \right)-\star_3dH_{\text{im}}\,,\qquad
 F^1=d\left( \frac{H_{\text{im}}\ma{H}_1^2}{\ma{S}^2}(dt+ \omega) \right)\,,\nonumber
\end{gather}
where
\begin{equation}
 \ma{S}=\sqrt{H_{\text{im}}\ma{H}_1^3-H_{\text{im}}^2 H_0^2}\,.
\end{equation}
With the choice (\ref{eq:flat_harm}) and (\ref{eq:flat_omega}), this can be explicitly written as
\begin{align}
 ds^2=&\frac{\Sigma^2}{\Delta} dt^2+\frac{\Sigma}{\Delta}\left[-a\sin^2\!\theta\left(2\,\widehat{k Q}\,r+\widehat{q Q} \right)+2\,\widehat{kq}\,(r^2+a^2)\cos\theta\right] dt d\varphi\nonumber\\
      &-\frac{\Delta}{\Sigma (r^2+a^2)}dr^2-\frac{\Delta}{\Sigma}d\theta^2
        \label{eq:mod2_case1_expl_metric} \\
      &+\left[\frac1{4\Delta}\!\left[-a\sin^2\!\theta\left(2\,\widehat{k Q}\,r+\widehat{q Q} \right)+2\,\widehat{kq}(r^2+a^2)\cos\theta\right]^2\! - \frac{\Delta}{\Sigma^2}(r^2+a^2)\sin^2\!\theta\right]
        d\varphi^2\,, \nonumber
\end{align}
\begin{align}
    A^0=&-\frac{\Sigma}{\Delta^2}\left(\Sigma k_0+q_0 r+Q_0 a \cos\theta\right)\left(\Sigma
               k_{\text{im}}+q_{\text{im}}r+Q_{\text{im}} a \cos\theta\right)^2 dt \nonumber \\
            & +\! \left[\frac{\Sigma}{2\Delta^2}\left(\Sigma k_0\!
               +\! q_0 r\!+\!Q_0 a \cos\theta\right)\left(\Sigma k_{\text{im}}\!+\!q_{\text{im}} r\!+\!Q_{\text{im}}
               a\cos\theta\right)^2\left(2\widehat{k Q}\,r+\widehat{q Q} \right) + Q_{\text{im}}r\right]\cdot
              \nonumber \\
            &\cdot\frac{a\sin^2\!\theta}{\Sigma}d\varphi
              -\left[\frac{\Sigma}{\Delta^2}\left(\Sigma k_0+q_0 r+Q_0 a \cos\theta\right)\left(\Sigma
              k_{\text{im}}+q_{\text{im}} r+Q_{\text{im}} a \cos\theta\right)^2\widehat{k q}+q_{\text{im}}
              \right]\cdot \nonumber \\
            &\cdot\frac{(r^2+a^2) \cos\theta}{\Sigma}d\varphi\,,
\end{align}
\begin{align}
    A^1=&\frac{\Sigma}{\Delta^2}\left(\Sigma k_{\text{im}}+q_{\text{im}} r+Q_{\text{im}} a
               \cos\theta\right)\left[\Sigma(t/\tilde t_1+k_1)+q_1 r+Q_1 a \cos\theta\right]^2\nonumber\\
            &\quad\cdot\left[dt-\frac{1}{2\Sigma}\left[\left(2\widehat{k Q}r+\widehat{q Q}\right)a
              \sin^2\!\theta+\widehat{k q}(r^2+a^2) \cos\theta\right]d\varphi\right]\,,
\end{align}
\begin{equation}
    Z=-\frac{\left[\Sigma(t/\tilde t_1+k_1)+q_1 r+Q_1 a \cos\theta\right]^2}{\left(\Sigma k_{\text{im}}
        +q_{\text{im}} r+Q_{\text{im}} a\cos\theta\right)\left(\Sigma k_0+q_0 r+Q_0 a
        \cos\theta\right)+i\Delta}\,,
\end{equation}
where
\begin{gather}
 \Delta=\left\{\left[\Sigma (t/\tilde t_1+k_1)+q_1 r+Q_1 a\cos\theta\right]^3\left[\Sigma k_{\text{im}}+
             q_{\text{im}} r+Q_{\text{im}} a\cos\theta\right]\right. \nonumber \\
             \left.-\left[\Sigma k_0+q_0 r+Q_0 a\cos\theta\right]^2\left[\Sigma k_{\text{im}}+q_{\text{im}}
             r+Q_{\text{im}} a\cos\theta\right]^2\right\}^\frac12 \nonumber \\
            \Sigma=r^2+a^2\cos^2\!\theta\,, \qquad \widehat{xy}=x_0y_{\text{im}}-x_{\text{im}}y_0\,.
\end{gather}
In the case of flat gauging, $C_1=0$ (which is inequivalent to $C_0=0$ for this model), the results of
section \ref{baseflat} are still valid provided one exchanges $0$ and $1$ indices everywhere. Redefining $H_1\rightarrow 3\, H_1$, the solution simplifies to
\begin{gather}
 ds^2=\ma{S}^{-1}(dt+\omega)^2- \ma{S} d\vec y^{\,2}\,,  \qquad
 Z=-\frac{H_1-i\,\ma{U}}{H_{\text{im}}}\,, \\
 F^0=-d\left(\frac{H_{\text{im}}}{\ma{U}^2}(dt + \omega)\right)\,, \qquad
 F^1=d\left(\frac{H_1}{\ma{U}^2}(dt+ \omega) \right)-\star_3dH_{\text{im}}\,, \nonumber
\end{gather}
with
\begin{equation}
 \ma{S}=H_{\text{im}}\,\ma{U}=H_{\text{im}}\sqrt{3 H_1^2-2\ma{H}_0H_{\text{im}}}\,.
\end{equation}
Since the potential vanishes for $C_1=0$, this is also a (non-supersymmetric) time-dependent
solution of ungauged supergravity.

The metric with the same harmonic functions and $\omega$ as before can again be written in the form
(\ref{eq:mod2_case1_expl_metric}), but where now
\begin{eqnarray}
\widehat{xy}&=&3(x_1y_{\text{im}}-x_{\text{im}}y_1)\,, \qquad
\Delta=\left[\Sigma k_{\text{im}}+q_{\text{im}} r+Q_{\text{im}}a\cos\theta\right]\tilde\Delta\,, \nonumber \\
\tilde\Delta&=&\left\{3\left[\Sigma\, k_1+q_1 r+Q_1 a\cos\theta\right]^2 -\left[\Sigma(t/t_0+k_0)+
q_0 r+Q_0 a\cos\theta\right]\cdot\right. \nonumber \\
&&\left.\cdot\left[\Sigma k_{\text{im}}+q_{\text{im}} r+Q_{\text{im}} a\cos\theta\right]\right\}^\frac12\,,
\end{eqnarray}
while the other fields read
\begin{align}
    A^0=&-\frac{\Sigma}{\tilde\Delta^2}\left(\Sigma k_{\text{im}}+q_{\text{im}} r+Q_{\text{im}} a
               \cos\theta\right)\cdot \nonumber \\
    &\qquad\cdot\left\{ dt+\frac1{2\Sigma} \left[\left(2\widehat{k Q}r+\widehat{q Q} \right)a
      \sin^2\!\theta-2\widehat{k q}(r^2+a^2) \cos\theta\right]d\varphi\right\}\,,
\end{align}
\begin{align}
    A^1=&\frac{\Sigma}{\tilde\Delta^2}\left(\Sigma k_1+q_1 r+Q_1 a \cos\theta\right) dt\nonumber\\
        &-\frac12 \left[\frac{\Sigma}{\tilde\Delta^2}\left(\Sigma k_1+q_1 r+Q_1 a \cos\theta\right)
          \left(2\widehat{k Q}r+\widehat{q Q} \right)-2 Q_{\text{im}} r\right]\frac{a\sin^2\!\theta}
          {\Sigma}d\varphi \nonumber \\
        &+\left[\frac{\Sigma}{\tilde\Delta^2}\left(\Sigma k_1+q_1 r+Q_1 a \cos\theta\right)
          \widehat{k q}-q_{\text{im}}\right]\frac{(r^2+a^2) \cos\theta}{\Sigma}d\varphi\,,
\end{align}
\begin{equation}
    Z=-\frac{\Sigma k_1+q_1 r+Q_1 a \cos\theta - i\tilde\Delta}{\Sigma k_{\text{im}}+
        q_{\text{im}} r+Q_{\text{im}} a \cos\theta}\,.
\end{equation}

\subsection{Spherical base space}
Using the results of section \ref{basesphere}, the complete solution can be written as
\begin{gather}
 ds^2=\ma{S}^{-1}(dt+\hat \omega)^2- \ma{S} ds_{\text{S}^3}^2\,, \qquad
 Z=-\frac{\ma{T}^1-i\ma{S}\tilde H_{\text{im}}}{\ma{T}^0+i\ma{S}H_{\text{im}}}\,, \\
 F^0=-d\left[\frac{\ma{T}^0}{\ma{S}^2}(dt+\hat\omega)\right]-\star_3dH_{\text{im}}\,, \qquad   
 F^1=d\left[\frac{\ma{T}^1}{\ma{S}^2} (dt+\hat\omega)\right]+\frac{t_1}{t_0}\star_3dH_{\text{im}}\,,
 \nonumber
\end{gather}
where
\begin{gather}
 \ma{S}=\sqrt{-\ma{H}_0\left(\ma{T}^0+\tilde H_{\text{im}}^3\right)+\ma{H}_1\left( \ma{T}^1-
 \frac{4}{27}H_{\text{im}}\ma{H}_1^2\right)}\,, \nonumber \\
 \ma{T}^0=\tilde H_{\text{im}}^3+H_{\text{im}}\tilde H_{\text{im}} \ma{H}_1+H_{\text{im}}^2\ma{H}_0\,,
 \qquad \ma{T}^1=\frac{4}{9}H_{\text{im}}\ma{H}_1^2+\frac{1}{3}\tilde H_{\text{im}}^2\ma{H}_1-
 H_{\text{im}}\tilde H_{\text{im}}\ma{H}_0\,,\nonumber\\
 \ma{H}_\Lambda=\frac{t}{t_\Lambda}+H_\Lambda\,,\qquad\tilde H_{\text{im}}=2 t_1-\frac{t_1}{t_0}
 H_{\text{im}}\,,
\end{gather}
and $\hat\omega$ satisfies (\ref{eq:spheredomega}).
An explicit solution can be obtained with harmonics of the form (\ref{eq:sph_harm_choice}), obeying the constraint 
(\ref{eq:sphere_harm_constraint}), and $\hat\omega$ given by (\ref{eq:sphere_omega}).

\subsection{Berger sphere}

Making use of the results of section \ref{baseberger} the complete solution can be written as
\begin{gather}
 ds^2=\ma{S}^{-1}(dt\pm\sin\mu\cos\mu\,\sigma_3^L t+\hat\omega)^2-\ma{S}  ds_3^2\,,  \qquad
 Z=-\frac{\ma{T}^1-i\ma{S}\tilde K_{\text{im}}}{\ma{T}^0+i\ma{S}K_{\text{im}}}\,, \\
 F^0=-d\left[ \frac{\ma{T}^0}{\ma{S}^2}(dt\pm\sin\mu\cos\mu\,\sigma_3^L t+\hat\omega) \right]-\star_3[dK_{\text{im}}\pm\sin\mu\cos\mu\,\sigma_3^L K_{\text{im}}]\,,\nonumber \\
 F^1=d\left[\frac{\ma{T}^1}{\ma{S}^2}(dt\pm\sin\mu\cos\mu\,\sigma_3^L t+\hat\omega) \right]-\star_3[d\tilde K_{\text{im}}\pm\sin\mu\cos\mu\,\sigma_3^L\tilde K_{\text{im}}]\,, \nonumber
\end{gather}
where
\begin{gather}
 \ma{S}=\sqrt{-\ma{K}_0\left(\ma{T}^0+\tilde K_{\text{im}}^3\right)+\ma{K}_1\left( \ma{T}^1-
 \frac{4}{27}K_{\text{im}}\ma{K}_1^2 \right)}\,,\nonumber\\
 \ma{T}^0=\tilde K_{\text{im}}^3+K_{\text{im}}\tilde K_{\text{im}}\ma{K}_1+K_{\text{im}}^2
 \ma{K}_0\,, \qquad
 \ma{T}^1=\frac{4}{9}K_{\text{im}}\ma{K}_1^2+\frac13\tilde K_{\text{im}}^2\ma{K}_1-K_{\text{im}}
 \tilde K_{\text{im}}\ma{K}_0\,, \nonumber \\
 \ma{K}_\Lambda=\frac{t}{t_\Lambda}+K_\Lambda\,, \qquad \tilde K_{\text{im}}= t_1\cos\mu-
 \frac{t_1}{t_0}K_{\text{im}}\,.
\end{gather}
Here the functions $K_0$ and $K_1$ satisfy equ.~(\ref{eq:berger_pseudoharm2}), $K_{\text{im}}$ obeys
(\ref{eq:berger_pseudoharm1}), and the time-independent one-form $\hat \omega$ is a solution of
(\ref{eq:berger_omega}).

\section{Conclusions\label{conclusion}}

In this paper, we used the results of \cite{Meessen:2009ma}, where all solutions to matter-coupled
fake $N=2$, $d=4$ gauged supergravity admitting covariantly constant spinors were classified,
to construct dynamical rotating black holes in an expanding FLRW universe. This was done for two
different prepotentials that are both truncations of the stu model and correspond to just one vector
multiplet. The cosmic expansion was thereby driven by two $\text{U}(1)$ gauge fields and by a complex
scalar that rolls down its potential. We considered three different choices for the Gauduchon-Tod
base space over which the four-dimensional geometry is fibered, namely flat space,
the three-sphere and the Berger sphere, and saw how the usual recipe in ungauged supergravity, where
extremal black holes are given in terms of harmonic functions on three-dimensional Euclidean space,
generalizes to a cosmological context. Some possible extensions and questions for future work are:
\begin{itemize}
\item Study more in detail the physics of the constructed solutions, for instance the presence
of trapping horizons \cite{Hayward:1993wb}, and see whether a first law of
trapping horizons \cite{Hayward:1997jp} holds.
\item Extend the analytic studies of nonrotating black hole collisions in de~Sitter space performed in
\cite{Kastor:1992nn,Brill:1993tm} to the more general solutions considered here, and see how the
results depend on the rotation, the cosmological scale factor different from dS, and the spatial
curvature of the underlying FLRW cosmology.
\end{itemize}
We hope to come back to these points in a future publication.

% \bibliography{papers}{}
% \bibliographystyle{JHEP}
\end{document}